# How water binds to microcline feldspar (001)

Giada Franceschi,[1*] Andrea Conti,[1] Luca Lezuo,[1] Rainer Abart,[2] Florian Mittendorfer,[1] Michael Schmid,[1] Ulrike Diebold[1]

[1]*Institute of Applied Physics, TU Wien, 1040 Vienna, Austria*
[2]*Department of Lithospheric Research, Universität Wien, 1090 Vienna, Austria*

*franceschi@iap.tuwien.ac.at

## Abstract

Microcline feldspar ($KAlSi_3O_8$) is a common mineral with important roles for Earth's ecological balance. It participates in the carbon, potassium, and water cycles, contributing to $CO_2$ sequestration, soil formation, and atmospheric ice nucleation. To understand the fundamentals of these processes, it is essential to establish microcline's surface atomic structure and its interaction with the omnipresent water molecules. This work presents atomic-scale results on microcline's lowest-energy surface and its interaction with water, combining ultrahigh vacuum investigations by non-contact atomic force microscopy and X-ray photoelectron spectroscopy with density functional theory calculations. An ordered array of hydroxyls bonded to silicon or aluminum readily forms on the cleaved surface at room temperature. The distinct proton affinities of these hydroxyls influence the arrangement and orientation of the first water molecules binding to the surface, holding potential implications for the subsequent condensation of water.

## Main text

Feldspars are tectosilicates made of corner-sharing $AlO_4$ and $SiO_4$ tetrahedra and varying ratios of Ca, Na, and K ions. They are ubiquitous and partake in maintaining our planet's delicate equilibrium. Feldspars largely compose the rocks we stand on, and are active at sequestrating atmospheric $CO_2$.[1] Through weathering processes, they transform into clays and create soils, providing essential nutrients for plant growth.[2] Furthermore, they exist as airborne dust particles in the atmosphere, where they influence ice nucleation (IN) and cloud formation, profoundly impacting global weather patterns.[3] While all these crucial processes occur on the surfaces of feldspars, the current knowledge about the atomic structure of feldspar surfaces – and how it may affect their interaction with the environment – stems largely from computational works. Experimentally, most information regarding surface processes of feldspars is inferred from either indirect or bulk measurements.

 The lack of detailed knowledge of the surface chemistry of feldspars is evident in current research on ice nucleation. K-feldspars ($KAlSi_3O_8$) and particularly the lowest-temperature polymorph known as microcline (Fig. 1) are exceptionally active ice-nucleating agents in the atmosphere.[4–10] Many theoretical studies have tried correlating surface chemistry and IN activity by investigating the atomic-scale interaction of "perfect" microcline surfaces with water. Ab-initio

DFT calculations have shown that ice-like structures can grow atop a non-ice-like, mediating water layer directly adsorbed on the lowest-energy (001) surface of microcline.[11] However, molecular dynamics studies have fallen short in replicating spontaneous IN on microcline's low-index facets, even at temperatures well below the freezing point of water.[12,13] On the experimental front, studies on IN have predominantly relied on the observations of macroscopic ice crystals, focusing on the potential role of macroscopic defects on microcline rather than its surface chemistry.[9,14–17] To bridge current theoretical and experimental studies, direct, atomic-scale investigations of pristine microcline surfaces and their interaction with water are needed. Such studies may shed light on microcline's ability to support hydrogen-bonded networks – an important factor for ice nucleation on other silicate minerals.[18]

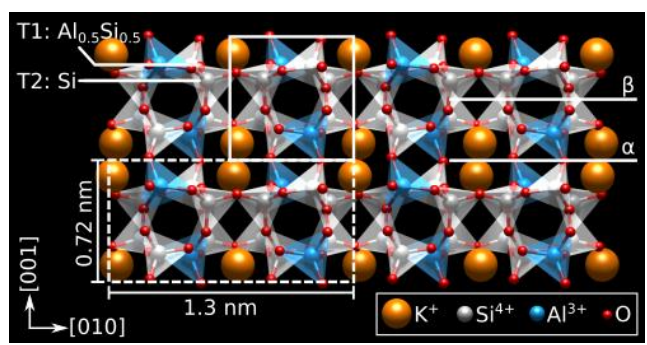

**Figure 1. DFT-optimized bulk structure of microcline feldspar.** View projected along the lowest-energy (001) plane. Polymerized $AlO_4^-$ and $SiO_4$ tetrahedra form a 3D network in the "mirror-crankshaft chain" configuration. The primitive unit cell (white, solid) contains 2 $AlO_4^-$ tetrahedra (blue), 2 $K^+$ ions (orange), and 6 $SiO_4$ tetrahedra (grey). The conventional unit cell (white, dashed), also used for the DFT calculations, contains double the atoms. Al ions sit exclusively at T1 sites, as opposed to higher-temperature feldspars where they can occupy both T1 and T2 sites. Marked on the right are the α and β cuts explored in this work.

Microcline's crystal structure is shown in Fig. 1. It is triclinic and centrosymmetric, comprising a 3D framework of corner-sharing $SiO_4$ and $AlO_4$ tetrahedra, with large cavities housing K ions. Cleaving along the (001) plane occurs easily; stacking along this direction comprises layers of K, mixed $SiO_4$ and $AlO_4$ tetrahedra, and $SiO_4$ tetrahedra. How microcline (001) is terminated after cleaving is debated.[13,19] Candidate cleaving planes are denoted as α and β in Fig. 1. Cleaving along plane α requires that half the number of bonds are broken as compared to plane β and should hence be favored. However, when hydroxylation is considered, plane β becomes more stable.[13] Note that the Al ions occupy only T1 sites in the tetrahedral framework. Following Löwenstein's rules for aluminosilicates, the Al ions that occupy 50% of the T1 sites will arrange in an ordered manner to minimize the overall electrostatic energy.[20] This makes microcline a "well-ordered" feldspar, in contrast to higher-temperature polymorphs where the Al ions are distributed among the T1 and T2 sites.[21]

This work aims to unveil the atomic structure of the cleaved (001) surface of microcline feldspar and its interaction with water under controlled conditions. Direct experimental investigations by atomically resolved non-contact atomic force microscopy (AFM) with a qPlus sensor[22] are complemented by X-ray photoelectron spectroscopy (XPS) and density functional theory (DFT) calculations. The measurements build on previous atomically resolved investigations of water structures adsorbed on ordered surfaces[23–30] but take the significant leap forward of tackling a large band gap material such as microcline. Microcline (001) is found to cleave at plane α, which readily hydroxylates even when cleaving in ultrahigh vacuum (UHV) at 300 K. The resulting surface hydroxyls (bonded to either Si or Al) are arranged in a buckled honeycomb pattern and template the adsorption of $H_2O$ molecules in an orderly fashion.

**The UHV-cleaved surface.** The (001)-oriented natural microcline mineral grain used for the present UHV study was characterized ex situ by microprobe analysis and photomicrography using (001)-oriented thin sections (Fig. S1). The sample is largely composed of microcline but also features small and sparse domains of Na-rich feldspar (albite), small quartz inclusions, and accessory hematite inclusions that give the mineral a reddish stain. Present are also sub-micrometer-sized inclusions of clay minerals, i.e., hydroxyl-bearing sheet silicates. The XPS data acquired on UHV-cleaved feldspar are in line with the ex-situ characterization. The survey and K $2p$ + C $1s$ region in Fig. S2 show that the cleaved surface is free of contaminants and features the expected elements (K, Si, Al, O), plus a minor contribution of Na, likely from the Na-rich feldspar regions.

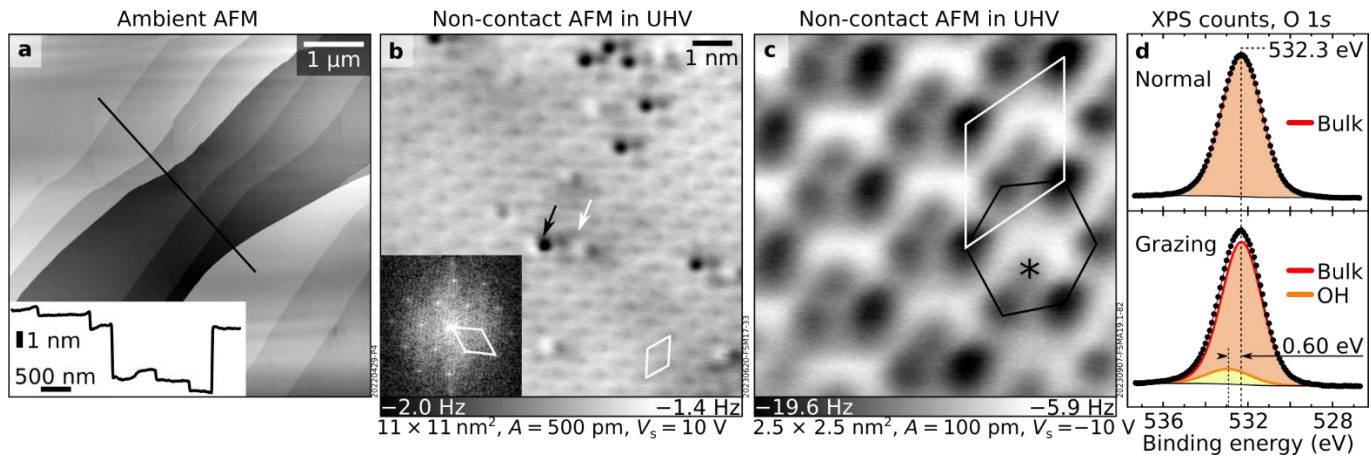

**Figure 2. Cleaved microcline feldspar (001).** (a) 6 × 6 μm² ambient AFM image. Inset: line profile taken at the solid line in the main panel. (b, c) Constant-height, non-contact AFM images of the UHV-cleaved surface. (b) Overview acquired with a $CuO_x$-terminated tip. Two types of point defects highlighted by arrows are visible on a regular lattice; inset: Fourier transform of (b). (c) Small-scale image acquired with a Cu-terminated tip. Highlighted are the honeycomb lattice with two sets of differently protruding features (black), a feature inside the honeycomb (asterisk), and the primitive unit cell (white). (d) O $1s$ core-level peaks in normal and grazing emission normalized to the area of the respective bulk components. The grazing-emission spectrum can best be fitted by adding a small contribution (yellow) at a binding energy higher than the main peak.

As expected from its known cleaving properties, the (001) microcline surface appears flat in ambient AFM images (Fig. 2a). Terraces are hundreds of nanometers in size and are separated by steps with heights that are multiples of the unit cell. Occasionally, areas with smaller terrace sizes were observed (Fig. S3). The surface also appears flat in AFM after UHV cleaving (Fig. 2b). It is well-ordered (see the Fourier transform in the inset), except for sparse bright and dark point defects. The appearance of the defect-free areas depends sensitively on the tip termination (Fig. S7) and relative tip-sample distance (Fig. S8). The sharpest tips (either Cu and CuO$_x$[31]) produce the contrast shown in Fig. 2c: a distorted honeycomb lattice (black) framed by two sets of differently attractive features, plus an additional feature inside the honeycomb (asterisk).

During cleaving, a water pressure burst was observed in the UHV chamber. The water may either derive from the clay inclusions in the microcline mineral grain or from micro- and nanometer-sized fluid inclusions that are typically associated with the interfaces between K-rich and Na-rich domains (see Section S2). While XPS cannot directly detect hydrogen, core-level shifts of elements that H may bind to, such as O, can be used to deduce the presence of water or hydroxyls on the surface. Figure 2d compares XPS O $1s$ peaks acquired on a UHV-cleaved surface in normal and grazing emission. The normal-emission peak, dominated by subsurface layers, is fit by one component (532.30 eV after binding energy correction for charging, see Section S1). The more surface-sensitive grazing-emission spectrum features a slightly shifted peak. Below, this is explained as an additional contribution due to surface OH groups at higher binding energy forming when water is released during cleaving.

**Dosing H$_2$O on the UHV-cleaved surface at low temperature.** The evolution of the surface upon dosing H$_2$O vapor at 100 K was followed both in XPS and AFM (Fig. 3). In grazing-emission XPS (Fig. 3a), a third component grows in the O $1s$ region. It is separated by 1.2 eV from the main component and is associated to molecular H$_2$O. In AFM (Figs. 3b−d), dark (attractive) features appear on the surface that gradually fill up a hexagonal lattice with the same periodicity as the cleaved surface. The tip can interact with these features and displace them to different lattice positions (in Fig. 3c, the circle highlights such an event; arrows indicate three water species displaced due to interaction with the tip). The attractive contrast in Fig. 3b was obtained with a Cu-terminated tip. Figure 5c, below, shows an image acquired with a CO-terminated tip, evidencing a bright (repulsive) contrast on the water species instead.

If the sample dosed with H$_2$O at 100 K is warmed to 300 K, the surface recovers the same appearance as an as-cleaved sample (Fig. S10), i.e., the features observed in Figs. 3b−d desorb from the surface. A desorption temperature between 150 K and 160 K was estimated from XPS (Section S1).

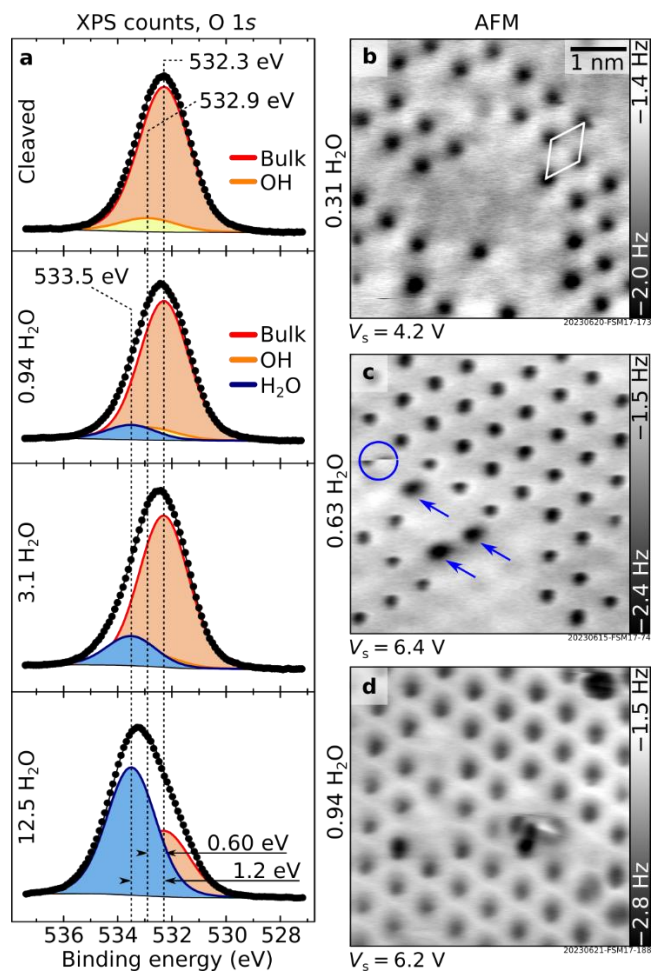

**Figure 3. Adsorption of H₂O at 100 K.** (a) Experimental data (dots) and fits (solid lines) of O 1*s* core-level peaks of UHV-cleaved microcline exposed to H₂O vapor at 100 K (Al Kα, pass energy 20 eV, 70° grazing emission; doses are expressed as the number of H₂O molecules per primitive unit cell dosed at 100 K.) (b−d) 6 × 6 nm² AFM images of microcline (001) exposed to H₂O vapor at 100 K. All images were acquired with a qPlus sensor and oscillation amplitude A = 500 pm; see Section S1 for $V_s$. In panel (b) the unit cell is highlighted in white. From one experiment to the next, the tip was slightly modified through interaction with water species. In panel (c), the circle highlights such an interaction event; arrows indicate three water species that have been displaced from their original lattice position due to the interaction with the tip.

**Computational results.** DFT calculations were performed for two different terminations of microcline (001), namely the α (between K planes) and the β (between Si-Si planes) cuts (Fig. 1). Calculations were performed on the dry as well as on water-exposed surfaces. The full set of calculations is discussed in detail in Section S2. Figures 4a−c focus on the results obtained on the α cut. Based on the phase diagram in Fig. S4a, plotting surface energies as a function of the water chemical potential, the α cut is the most stable termination in a wide range of experimental conditions: At a temperature of 300 K, it has the lowest energy across pressures ranging from UHV to ambient pressure.

As seen from Fig. 4a, the dry α cut is essentially bulk truncated. Cleaving breaks the surface O-Al bonds, leaving O atoms on the topmost Si atoms and producing undercoordinated surface Al. (Fig. S5 shows that breaking O-Si or mixed O-Al and O-Si is less favorable.) Water readily dissociates on this termination (see Fig. 4b), with an adsorption energy of −3.3 eV/H$_2$O. The first H$_2$O molecule per primitive surface unit cell (u.c.) splits without a barrier, donating one proton to the Si-backbonded surface O atom and the split-off OH to the undercoordinated Al ion (Fig. 4b), i.e., creating a silanol and aluminol species. A coverage of one H$_2$O molecule per u.c. is enough to fully hydroxylate the surface. The large adsorption energy explains why the surface remained protonated during molecular dynamics simulations with large quantities of water.[13] To explore how additional water adsorbs on the fully hydroxylated α surface, calculations were run with one extra H$_2$O molecule per u.c.. Consistent with computational results[11] and the XPS data in Fig. 3a, the additional molecule remains undissociated. As seen from Fig. 4c, it accepts a hydrogen bond from Si-OH and donates one to Al-OH. In agreement with Ref. [11], the molecule adsorbs with ≈−0.8 eV binding energy.

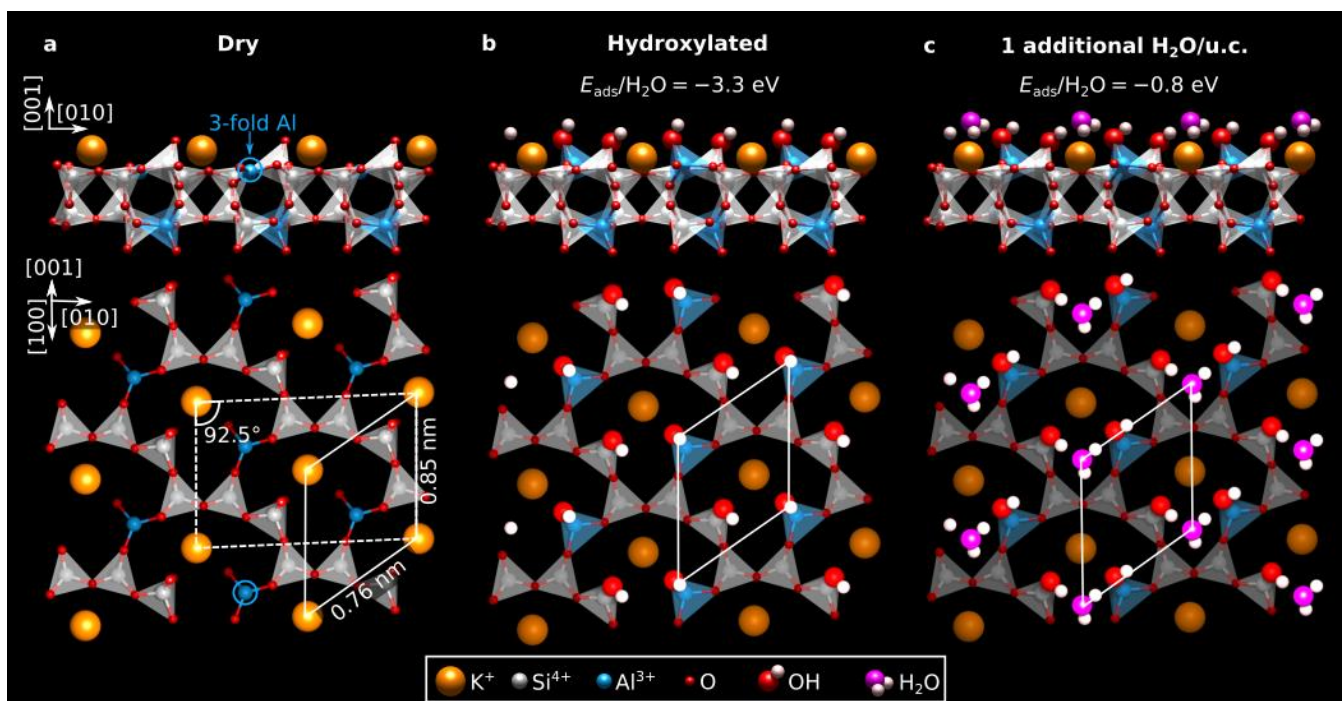

**Figure 4. DFT-relaxed α cuts of microcline feldspar (001).** (a) No water adsorbed. (b) 1 dissociated H$_2$O/u.c. (c) 1 additional H$_2$O molecule/u.c. over the hydroxylated surface. Relative adsorption energies are reported above the corresponding models. The conventional and primitive unit cells are marked with dashed and solid lines, respectively.

Figure 5 compares experimental AFM images of the cleaved and water-dosed surface with AFM simulations from the theoretical models of Figs. 4b, c. Simulated images of the hydroxylated α cut reproduce the AFM contrast on the cleaved surface (see Figs. 5a, b, obtained with CuO$_x$ tips; Fig. S8 shows results obtained Cu-terminated tips instead). Both CuO$_x$ (Fig. 5a) and Cu tips (Fig. 2c) show a honeycomb pattern. Each honeycomb is composed of two sets of species with different contrast, highlighted by the black and white circles triangles in Fig. 5a. The darker set (stronger attractive interaction of the AFM tip; marked by black circles) is assigned to the most protruding Al-OH, the fainter (white circles) to Si-OH. A faint feature is observed inside the honeycomb and is marked by an asterisk (this is more evident with sharper tips, see Fig. 2c). Based on the correspondence with the DFT relaxed model, it is assigned to the highest-lying K ion. Note that microcline has a centrosymmetric crystal structure. Hence, the $(001)$ and $(00\bar{1})$ terminations should be mirror symmetric. Consistently, mirror-symmetric AFM simulations and experimental images are obtained on opposite terminations (Section S4).

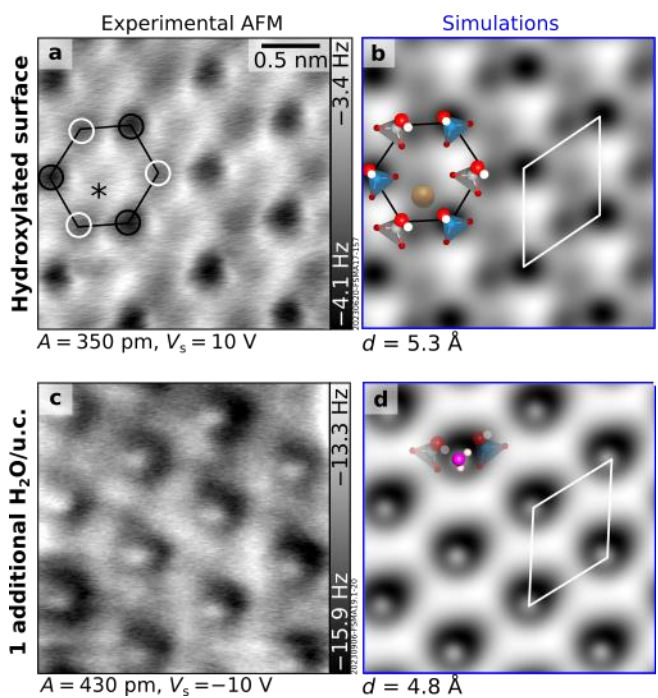

**Figure 5. Comparison between experimental and simulated AFM images.** Black frames, left) and simulated (blue frames, right) AFM images of the (a, b) hydroxylated microcline (001) α surface and (c, d) the same surface after dosing one additional water molecule per primitive unit cell (u.c.) at 100 K. All images are $2.5 \times 2.5$ nm². White rhombi identify the primitive unit cell. In panel (a), black and white circles mark hydroxyls bound to Al and Si, respectively. The asterisk marks the feature assigned to K. AFM simulations were performed at the tip-sample distances d noted below the corresponding images. Different tips were used in both experiments and simulations: (a, b) CuOx-terminated; (c, d) CO-terminated. Figure S8 shows experimental and simulated images of both surfaces with Cu tips.

A good match is also obtained between the experimental AFM image of the cleaved surface dosed with one $H_2O$ molecule per unit cell at 100 K (Fig. 5c) and the simulation obtained from the model of Fig. 4c, i.e., one additional $H_2O$ per unit cell on top of the hydroxylated α cut (Fig. 5d). Both show a hexagonal pattern of protruding features with the same unit cell as the hydroxylated surface. These features are imaged in the repulsive regime (bright) with a CO-terminated tip, and in the attractive regime (dark) with a Cu-terminated tip (Fig. S8).

The computational data presented in Section S3 show that the α cut is more stable than the β cut under UHV conditions, i.e., the sample should cleave between the K planes and relax to a quasi-bulk-truncated termination. If sufficient water is available in UHV, this termination should readily hydroxylate due to the large adsorption energy of $H_2O$, as also evident from the phase diagram of Fig. S4a. Previous literature reported that the hydroxylated β cut should be more stable than the hydroxylated α cut in UHV at 0 K.[13] However, this situation will not be obtained experimentally. The sample will cleave at the energetically preferred plane (the α cut, where the least number of bonds are broken). If enough water is available, the α cut will become hydroxylated. Kinetics at room temperature is insufficient to switch to the hydroxylated β plane.

All evidence suggests that microcline (001) cleaves at the α cut and readily hydroxylates in UHV at 300 K – even without any intentional water supply. While one could consider that microcline cleaves preferentially at 'special', hydroxylated planes, the absence of step-bunching (see line profile in Fig. 2a) speaks against this hypothesis. The water needed for hydroxylation is likely provided by clay or fluid inclusions in the natural minerals (see Section S2) exposed during the cleaving procedure. Based on the DFT-predicted adsorption energies, any available water molecules will stick with 100% probability on the microcline surface and dissociate without a barrier to form two hydroxyls. Based on mass-spectrometer measurements, the amount released through cleaving suffices for full hydroxylation (see Section S1 and Fig. S10c). While it is somewhat surprising that the surfaces are immediately hydroxylated and 'dry' surfaces are not produced even in the most pristine UHV environment, the energetics (see the phase diagram of Fig. S4) suggest that the resulting, fully hydroxylated surfaces will also be present in ambient conditions.

The surface OH groups after cleaving are evidenced by a small component at higher binding energy in the grazing-emission O 1$s$ spectrum (Fig. 2b). This signal sits between the main O 1$s$ component (532.30 eV) and the molecular $H_2O$ component obtained by dosing $H_2O$ at 100 K (533.5 eV, see Fig. 3a), as typical for OH on other water-exposed oxides.[26,29,32] The energy differences between the bulk O 1$s$ peak and the ones assigned to OH and $H_2O$ peaks (0.60 eV and 1.20 eV, respectively) are reasonably reproduced by DFT calculations: Core-level-energy shifts of ≈0.5 eV and ≈0.9 eV are predicted for OH and $H_2O$ (details in Section S1). The presence of hydroxyls after cleaving is also supported by the identical appearance of the surface after dosing $H_2O$ at 100 K followed by warm-up to 300 K (Fig. S10). Based on the strong adsorption energies of −3.3 eV predicted by DFT, the hydroxyls are expected to remain on the surface at 300 K. Finally, DFT predicts adsorption energies of −0.8 eV for $H_2O$ molecules adsorbed on the hydroxylated surface (Fig. 4c), i.e., temperatures lower than 300 K will be needed for adsorption. Consistently, XPS shows that molecular $H_2O$ starts desorbing between 150 K and 160 K, corresponding to an adsorption energy of ≈ −0.6 eV.[33] The picture is validated by the good match between the experimental images and the simulations from the DFT-relaxed models (Fig. 5).

The different types of hydroxyls found at the microcline surface (Al-OH and Si-OH) should affect the anchoring of subsequent $H_2O$ molecules. The bond between Al and OH is weaker than the one between Si and OH due to the smaller charge of Al (3+) compared to Si (4+). As a result, the proton bound to Si-O should be released more easily than the one bound to Al-O; in other words, Si-OH should be more acidic than Al-OH. As seen from Fig. 4c, such a difference in acidity influences the adsorption configuration of additional $H_2O$ molecules on the hydroxylated surface. As expected, the more acidic Si-OH donates a proton to the $H_2O$ molecule, while Al-OH accepts it. The model in Fig. 4c is further supported by the good match between experimental and simulated images seen in Fig 5.

That OH sites are important for stabilizing water molecules should not surprise. Under ambient conditions, hydroxyls exposed at oxide surfaces participate to the formation of wetting layers.[34,35] At lower temperatures, the OH surface density is a good predictor of IN abilities.[36] OH sites and the H bonds they offer appear to be more important than the electrostatic interaction with surface $K^+$ ions. The latter remain snug in their position, contrary to what happens upon immersion in liquid, where K ions are readily exchanged for protons.[19,37] On the other hand, when there is no opportunity for surface OH groups, templating an ordered array of $H_2O$ molecules may be challenging. Muscovite mica – another K-rich aluminosilicate of composition $KAl_2(Si_3Al)O_{10}(OH)_2$ – exemplifies this. When cleaved in UHV, muscovite exposes undercoordinated K ions lying on an otherwise bulk-truncated surface.[38] Water dosed at 100 K in UHV on this system adsorbs molecularly rather than dissociatively, completing the hydration shell of the surface cations and triggering the formation of 3D clusters rather than an ordered network of $H_2O$ molecules.[39]

It is interesting to compare the presented study with ice nucleation experiments performed on macroscopic microcline crystals. In this work, XPS shows that $H_2O$ molecules dosed at 100 K disappear between 150 and 160 K in isobaric equilibrium measurements at a partial pressure of $1.5 \times 10^{-8}$ mbar. This corresponds to a chemical potential of water $\mu_{H_2O}$ between $-0.57$ eV and $-0.53$ eV (see Section S1). These values are aligned with existing immersion-freezing[40] and deposition-mode experiments[9] on microcline, where ice condenses at $\mu_{H_2O}$ values between $-0.54$ eV and $-0.55$ eV. The matching values of the water chemical potential indicate that the conditions at which ice nucleates in the two cases are comparable. However, this alone is not enough to draw conclusions about the mechanism underlying ice nucleation on microcline. Macroscopic defects[9,14–17,41] are known to play an important role for IN, but the circumstances leading to IN are not clear.[9] A comparison of the IN activity of the same feldspar surfaces in immersion freezing and deposition modes showed that these provided two poorly correlated sets of active sites. A handful of sites though were active in both modes, pointing to a common nucleation mechanism.[9] Interestingly, crystalline ice structures with the same epitaxial orientation were observed in both modes, a potential evidence that nucleation occurs on surface features of the crystalline substrate rather than on contaminants.[9] The importance of the surface chemistry of crystalline phases is supported by the observed decreased IN efficiencies on amorphous silicates compared to crystalline ones.[42–44] It is possible that the ordered anchoring of $H_2O$ molecules observed under UHV conditions offers the opportunity to later create H-bonded water layers. In turn, this may relate to the observed crystalline ice structure. At this stage, however, it is premature to draw definite conclusions about the relative importance of surface chemistry vs. surface defects for ice nucleation on microcline.

The specific superiority of microcline compared to other K-feldspars also remains up for debate. Assuming that all K-feldspars have comparable macroscopic defects, the differences in their IN activities must relate to their intrinsic surface chemistry. By analogy with microcline, all (001)-oriented K-feldspars should cleave at the α cut and readily hydroxylate when exposed to small quantities of water. The main difference between microcline and other polymorphs will be the number (smaller) and arrangement (more disordered) of the surface Al ions and, consequently, aluminol groups. As mentioned above, microcline is the most-ordered K-feldspar, with Al ions occupying only the surface T1 sites; in other feldspars, Al ions are randomly distributed in surface T1 and subsurface T2 sites, see Fig. 1. Since aluminol and silanol groups have different binding strengths and proton affinities,[11] additional $H_2O$ molecules landing on disordered K-feldspars will find inequivalent, disordered, binding sites. This might disrupt the creation of an ordered first $H_2O$ adlayer, muddling the adsorption of additional water and decreasing overall IN abilities.

In summary, this study combines UHV analyses by AFM and XPS with DFT calculations to investigate the atomic-scale details of microcline feldspar (001) and their interaction with water. The UHV-cleaved surface strongly reacts with water at room temperature, producing Si- and Al-bonded hydroxyls visible as a buckled honeycomb pattern in atomically resolved AFM images. The different acidity of the long-range-ordered aluminol and silanol groups enforces a specific adsorption configuration for $H_2O$ molecules on this surface, carrying potential implications for the subsequent condensation of water molecules.

**Supporting Information**. Section S1: Methods (UHV setup and characterization, ex-situ characterization, computational details). Section S2: Further characterization of microcline feldspar (thin-section characterization, XPS, cleaving procedures, further ambient AFM images, optical approach in UHV). Section S3: Additional computational results (additional details about the β cut, additional details about the α cut, performance of r²SCAN and r²SCAN-D3 functionals compared, phase diagram as a function of the water chemical potential). Section S4: Considerations about symmetry. Section S5: Additional experiments and simulated images. Section S6: Δf-z curves. Section S7: Arguments for the ready hydroxylation of the as-cleaved surface. Section S8: Imaging in the presence of surface charges.

## Author Contributions

Conceptualization: GF, UD. Investigation: GF, AC, LL, RA. Supervision: GF, FM, UD. Validation: MS, UD. Writing—original draft: GF. Writing—review & editing: all authors.

## Acknowledgements

GF, AC, LL, and UD acknowledge support from the European Research Council (ERC) under the European Union's Horizon 2020 research and innovation programme (grant agreement No. 883395, Advanced Research Grant 'WatFun'). The computational results have been achieved using the Vienna Scientific Cluster (VSC). Prof. Uwe Kolitsch from the Natural History Museum of Vienna is acknowledged for providing the samples used for this work, and Prof. Gerald Giester for determining the orientation of the sample used in this work by X-ray Diffraction. Discussions with Prof. Angelika Kühnle, Dr. Pablo Piaggi, and Prof. Annabella Selloni are gratefully acknowledged.

Supplementary Information for

# How water binds to microcline feldspar (001)

Giada Franceschi* *et al.*

*franceschi@iap.tuwien.ac.at

**This file includes:**

Supplementary Text

## Supplementary Text

### Section S1: Methods

**UHV setup and characterization.** The experiments were carried out in a UHV setup consisting of two interconnected chambers: A preparation chamber for sample cleaving and XPS measurements and an adjacent chamber for AFM measurements (base pressures below $1 \times 10^{-10}$ mbar and $2 \times 10^{-11}$ mbar, respectively).

Natural microcline feldspar from Russia obtained by kind concession of Prof. Uwe Kolitsch from the Natural Museum of History of Vienna were characterized ex situ before the UHV measurements (see Section S2, reporting results from photomicrography, electron probe microanalyzer, and ambient AFM). The samples were glued on Omicron-style stainless steel sample plates with UHV-compatible epoxy glue (EPO-TEK T7110-38). They were cleaved in UHV at room temperature before each experiment with two methods shown in Figs. S3a, b, which yielded consistent results: (i) by using a wobble stick to apply a tangential force to a metal stud glued on top of the sample.[45] The sample portion initially covered by the stud is thus cleaved and ready for XPS and AFM investigations; (ii) by using a guillotine-type cleaver[46] that shaved off the top part of the sample.

To study water adsorption on the microcline surface at low temperatures, water vapor was dosed from a leak valve while keeping the sample holder on the manipulator of the preparation chamber to 100 K with liquid nitrogen. The amount of water dosed on the surface is always expressed as the number of $H_2O$ molecules per primitive unit cell (u.c.). The calibration was performed based on the dosing (water partial pressures × time) needed to obtain a coverage of 1 $H_2O$/u.c. at 100 K (assuming 100% sticking probability). This corresponds to 32 s at $1.3 \times 10^{-8}$ mbar, in turn equal to 0.32 Langmuir (L, where 1 L is defined by an exposure time of 1 s at $1.3 \times 10^{-6}$ mbar).

With each cleave, the partial pressure of water in the UHV chamber increased to $\approx 2 \times 10^{-7}$ mbar for a few seconds, for a total dose of $\approx 24$ Langmuir (see Fig. S10c and related discussion). Since the hydroxylated surface requires a nominal dose of 0.32 L (the hydroxylated surface has the same density of $H_2O$ molecules as the water-dosed surface), the observed 'pressure burst' should be enough to hydroxylate the surface fully.

XPS was performed with a non-monochromatic dual-anode Mg/Al X-ray source (SPECS XR 50) and a hemispherical analyzer (SPECS Phoibos 100). Spectra were acquired in normal and grazing emission (70° from the surface normal). The intensities and positions of the Al-Kα-excited XPS peaks were evaluated with CasaXPS after subtracting a Shirley-type background. Due to the insulating nature of the samples (bandgap of 7.6−7.7 eV [47]), the XPS spectra showed shifts to apparent higher binding energies (between 5 and 7 eV). The magnitude of the shift depends on the amount and type of surface adsorbates, XPS acquisition geometry, and sample thickness. For the display and analysis of the XPS data, an energy correction was applied to all spectra: The Si 2*p* core-level peak was set to 103.10 eV, as reported in the literature.[48] This resulted in an O $1s$ peak at a binding energy of 532.30 eV at normal emission.

Table 1 summarizes the constraints applied to the fits. The O $1s$ peak of the cleaved surface was fit by comparing normal and grazing emission acquisitions. The normal-emission spectrum was fit by component 1 alone. Fitting the grazing-emission spectrum additionally required component 2. In the main text, component 2 is assigned to surface OH species that saturate the cleaved surface at room temperature. Increasing amounts of $H_2O$ at 100 K induced the growth of a third component (assigned to molecular $H_2O$, see main text). Its position and FWHM were determined from high-dose experiments (>4 L, or 12.5 $H_2O$/u.c.), which were then constrained to fit the lower doses. For the fits of the low-temperature water experiments, the intensity ratio of components 1 and 2 was constrained to the value found on the cleaved surface, under the assumption that molecular $H_2O$ grows onto the fully hydroxylated surface (see main text).

**Table 1. Details about the XPS fitting components of Fig. 2.** The shape (LA=asymmetric Lorentzian), full-width half maximum, and position were constrained for all peaks.

|  | Identifier | Shape | FWHM | Position (eV) | Area |
|---|---|---|---|---|---|
| O $1s$ 1 | O $1s$ cleaved | LA(1.53,243) | 2.24 | 532.30 | Free |
| O $1s$ 2 | OH cleaved | LA(1.53,243) | 2.5 | (O $1s$ 1) + 0.60 | (Area O $1s$ 1) ×0.117 (for molecular $H_2O$ dosing) |
| O $1s$ 3 | $H_2O$ | LA(1.43,243) | 2.2 | (O $1s$ 1) + 1.20 | Free |

XPS was used to obtain an approximate isobar for molecular $H_2O$. Here 2 L $H_2O$ were dosed at 100 K. In a water background pressure of $\approx 1.5 \times 10^{-8}$ mbar, the sample was warmed to increasingly higher temperatures in steps of 10 K. XPS spectra (O $1s$, K $2p$, and Si $2p$ for energy correction) were acquired at each stage. To reach a situation as close to thermodynamic equilibrium as possible, the sample was kept in the water background for $\approx 20$ min at each temperature before each measurement. The temperature where the coverage of the molecular water roughly halved was between 150 K and 160 K.

The AFM measurements were performed at 4.7 K using a commercial Omicron qPlus LT head and a differential cryogenic amplifier.[49] Frequency-modulated non-contact AFM mode was used. The tuning-fork-based AFM sensors ($k \approx 2,000-3,500$ N/m, $f_0 \approx 32$ kHz, Q $\approx 50,000$) had a separate contact for tunneling current. The electrochemically etched W tips were cleaned by field emission.[50] Before each measurement, the tips were prepared on an oxygen-exposed Cu(110) single crystal by repeated indentation and voltage pulses. Cu-, CuO$_x$-, and CO-terminated tips were prepared on the oxygen-induced reconstruction of Cu(110)[31] to exhibit a frequency shift smaller (in absolute value) than $-1.5$ Hz. They were used to image the cleaved and water-exposed microcline surface. At times, the tip interacted with point defects (strongly attractive features, likely adsorbates) present on the surface, which easily snapped to the tip. The tip-to-sample approach was performed carefully to avoid the risk of crashing into the insulating, possibly charged, microcline surface. First, the tip was manually brought closer to a flat area of the surface, as judged by looking from an optical camera (Fig. S2d). This was followed by an automatic approach with a setpoint of −800 mHz, which ensures that the approach is stopped well before reaching the surface. Then, the controller was switched off and the tip was gradually approached in constant-height mode until an AFM contrast was visible while scanning in x and y, at which point potential tilts of the surface were corrected for. All the AFM images presented here were acquired in constant-height mode.

Like other insulators, the as-cleaved samples exhibit surface charges that make AFM measurements difficult.[38,51] The charge could be effectively remediated by irradiating the cleaved surface with X-rays from our XPS setup for one minute. This treatment did not introduce any spurious contaminations. Residual fields can be compensated by applying a bias voltage between the tip and sample. All measurements were performed by applying a bias voltage $V_s$ to minimize the electric field between the tip and the sample, as judged from local contact potential difference (LCPD) measurements by the Kelvin parabola method.[52] In other words, the bias voltage was set to the maximum of the LCPD parabola. These $V_s$ are reported in the images and correspond to the bias applied to the back of the sample plate while keeping the potential of the tip close to ground. When the surface was not irradiated by X-rays before the AFM measurements, it was not possible to compensate the potential with the maximum ±10 V voltage range delivered by the microscope controller, resulting in large absolute values of frequency shifts in the constant-height AFM images (Fig. S11 exemplifies this; background modulations due to surface charging are evident). Frequency shift vs. tip-sample distance curves were acquired on representative features on the cleaved and water-dosed surfaces as discussed in Section S6.

**Ex-situ characterization.** Ambient AFM images were acquired in air with an Agilent 5500 ambient AFM in intermittent contact mode with Si tips on Si cantilevers. X-ray diffraction (XRD) was used to determine the orientation of the sample using a small chip from a centimeter-sized feldspar crystal, keeping track of the mutual orientation. The chip was mounted on a Nonius KappaCCD 4-circle diffractometer run with Mo radiation. The acquisition of ten frames was sufficient to unambiguously identify the crystal orientation of the chip and of the larger feldspar crystal. For photomicrography and electron probe microanalysis (EMPA), a slab of roughly two millimeter thickness was cut from the main feldspar crystal parallel to (001) using a diamond wire saw. The slab was embedded into a stub of epoxy resin. The surface was ground and subsequently polished using diamond paste down to a grain size of 0.25 μm to obtain a smooth plane surface needed for mineral chemical analysis with EPMA. The polished rock chip had a thickness of about 500 μm and was translucent, allowing for polarization microscopy in transmitted light. Polarization microscopy was done on a Leica DM 4500 P polarization microscope with a CCD camera. For electron microscopy, the surface was carbon-coated to ensure electrical conductivity. Backscattered electron (BSE) images and mineral chemical analyses were done on a CAMECA SX Five EMPA equipped with a field emission electron source and five wavelength dispersive crystal spectrometers as well as an energy dispersive system for elemental analysis (accelerating voltage 15 kV, beam current 20 nA). Natural mineral and synthetic oxide standards were used to calibrate quantitative mineral chemical analyses.

**Computational methods (DFT).** DFT calculations were performed with the Vienna Ab-initio Simulation Package (VASP)[53,54] using the r$^2$SCAN-D3 [55] metaGGA exchange-correlation functional. This functional describes well the bulk structural properties; the lattice constants and angles deviate less than 0.4% from experimental values [56]. A comparison of selected values using the r$^2$SCAN functional[57] can be found in Section S3.

The bulk structure (Fig. 1) was optimized with a cutoff energy of 700 eV. A k-point mesh of $3 \times 2 \times 3$ was used to integrate the Brillouin zone. The unit cell used for the calculations is the conventional cell used in the literature (dashed in Fig. 1, with the following optimized lattice parameters: $a = 8.54$ Å, $b = 12.95$ Å, $c = 7.21$ Å, $\alpha = 90.65°$, $\beta = 116.17°$, $\gamma = 87.61°$). This is larger than the primitive unit cell (solid, Fig. 1). For studying the surfaces, the slabs were symmetric, made of 16 layers (i.e., 16 formula units of KAlSi$_3$O$_8$ when using the primitive unit cell in $x$ and $y$), and separated by 20 Å vacuum regions, unless otherwise specified. All atoms were free to relax. The surface calculations had a cutoff energy of 400 eV and a $3 \times 2 \times 1$ k-point mesh. Geometries were optimized using the conjugate gradient method. The structures were relaxed until achieving residual forces on the atoms smaller than 0.01 eV/Å and an energy convergence of $10^{-6}$ eV.

To determine the most stable hydroxylated structures, various starting configurations with distinct OH orientations were relaxed. Several configurations were also tested for the water molecules adsorbed on the hydroxylated surface (see Section S3).

The AFM images were simulated with the Probe Particle Model,[58,59] which includes Hartree-potential electrostatics and Lennard-Jones potentials as well as the elastic properties of the tip based on the methods described in Refs. [58,59]. CuO$_x$ and Cu tips were simulated with the following values of lateral and vertical spring constants and charges (CuO$_x$: $k_{x,y}$=161.9 N/m, $k_z$=271.1 N/m, effective tip charge of −0.05e; CO: $k_{x,y}$=1.7 N/m, $k_z$=326.9 N/m, effective tip charge of −0.005e; Cu: $k_{x,y}$=7.8 N/m, $k_z$=50.7 N/m, effective tip charge of −0.05e). Note that with values larger than 5.0 N/m, the tip's stiffness had only a minor influence on the appearance of the simulated images. The oscillation amplitude for each simulation (250−500 pm) always matched the one used in the corresponding experimental image. The best-fitting images were selected within tip-sample distances fitting the experimental ones; distances are always referenced to the most protruding surface atom.

The chemical potential of water used in the phase diagram of Fig. S4a is defined as $\mu_{H_2O}(T,p) = \mu_{H_2O}(T,p^o) + kT \ln\left(\frac{p}{p^o}\right)$. It provides the temperature and pressure dependence given the temperature dependence of $\mu_{H_2O}(T,p^o)$ at a particular pressure $p^o$ .[60] The reference state was chosen as the total energy $E_{H_2O}^{gas}$ of an isolated H$_2$O molecule in the gas phase. Assuming this reference and based on the number of water molecules $N_w$ and formula units of bulk feldspar $N_b$, the surface energies of symmetric slabs of area $A$ were calculated as $\gamma = \frac{E_{slab} - N_w E_{H_2O}^{gas} - N_b E_{bulk}^{feldspar}}{2A}$.

XPS core-level shifts were determined both in the initial and in the final state approximations.[61,62] The shifts of the OH and H$_2$O components with respect to the bulk O 1$s$ component were calculated as (0.46 eV, 0.87 eV) and as (0.51 eV, 0.95 eV) with the two approximations, respectively.

## Section S3: Further characterization of microcline feldspar

Thin-section characterization

Figure S1 displays a backscattered electron (BSE) image and a photomicrograph of a (001) oriented about 500 μm thick section of the same microcline grain as used for the AFM experiments. Most of the sample (bright areas in the BSE image of Fig. S1a) is identified as microcline based on the composition measured by electron probe micro analyser (EPMA) as $K_{0.94}Na_{0.06}Al_{1.01}Si_{2.99}O_8$ and the typical "microcline grid" appearance in the photomicrograph of Fig. S1b (see below). In the BSE images, two types of dark grey domains are visible. Thin, nearly vertically oriented dark grey features correspond to exsolution lamellae of Na-rich alkali feldspar (albite, $NaAlSi_3O_8$), which typically form by diffusion controlled solid-state reaction from Na-bearing K-feldspar during slow cooling from their primary crystallization temperatures.[1] The larger, irregularly shaped dark grey patches and vein-like structures are precipitates of albite, which, based on their appearance, are interpreted as having partially replaced the original K-rich feldspar in a fluid-mediated process during late-stage hydrothermal overprint. Typically, fluid-filled micro- and nano-porosity is concentrated at the interfaces between the original K-feldspar and the albite-rich precipitates.[2,3] In addition, small blebs of quartz were identified by EPMA measurements, which also appear dark grey on BSE images but are too small to be discerned in Fig. S1a. The chemical formula of the sodium-rich phases reads $Na_{0.95}K_{0.01}Ca_{0.04}Al_{1.04}Si_{2.96}O_8$, which is almost pure albite with about 4 mole % anorthite ($CaAl_2Si_2O_8$) component.

The photomicrograph of Fig. S1b was acquired in crossed polarized transmitted light. Note, as the specimen is about 500 μm thick, the color shades correspond to high-order interference colors. Color contrasts indicate differences in crystal orientation. The image evidences two sets of lamellae parallel to the traces of (010) and (100) planes. The lamellae correspond to albite twins extending parallel to (010) and pericline twins extending parallel to (100). Combined albite and pericline twining yields the characteristic "microcline grid", a diagnostic feature of this mineral species. In addition, darker, "cloudy" areas are visible. These likely correspond to "filling" by sub-micron sized inclusions of clay minerals, a feature that is frequently observed in potassium-rich alkali feldspars and is due to late stage hydrothermal processes.

XPS

The XPS survey in Fig. S2 shows that the UHV-cleaved microcline features all the expected elements (K, Si, Al, O), plus a contribution of Na. As discussed above, this arises from small and sparse albite domains present in the natural mineral. As shown by the K $2p$ + C $1s$ region acquired in grazing emission, the surface is free from so-called "adventitious" carbon.

K $2p$ spectra acquired in normal and grazing emission (insets) are fit by the same contributions and look essentially identical. This is different from the K-terminated surface of muscovite mica, where surface and bulk contributions show a core-level shift of 1.22 eV. In mica, these contributions cause different peak shapes in normal- and grazing-emission spectra.[4] On the other hand, for the hydroxylated α cut of microcline, DFT predicts a core-level shift of 0.13 eV between K $2p$ surface and bulk contributions, which cannot be resolved experimentally. This is consistent with the identical appearance of the K $2p$ spectra in normal and grazing emission. The smaller core-level shifts in microcline than in muscovite mica are likely due to the different types of surface K ions in the two cases: on microcline, the K ions are "embedded" in the surface (see Fig. 1 in the main text); on muscovite, the K ions significantly protrude over the surface, resembling isolated adatoms.

## Section S3: Additional computational results

Figure S4 summarizes the main outcomes of the DFT calculations performed on the α and β cuts of microcline (001). As mentioned in the main text, Al ions exclusively occupy T1 sites in the tetrahedral framework of bulk microcline[5] (see Fig. 1). Previous theoretical works have predicted that the hydroxylated surface would favor the occupation of T2 sites instead.[6] However, at room temperature, Al mobility is insufficient for them to go from occupying the bulk equilibrium sites (T1) in the tetrahedral network[5] to the less stable T2 sites. Hence, the calculations performed here maintain the occupation of the Al ions on the T1 sites.

Several calculations were tested for the hydroxylated surfaces and for the water-dosed hydroxylated surfaces. For the hydroxylated surfaces, five starting configurations with distinct OH orientations were relaxed, specifically: OH pointing vertically, and along [100], [$\bar{1}$00], [010], and [0$\bar{1}$0]. The most stable configuration found for the α cut (Fig. 4b of the main text and Fig. S4c) can be obtained by initially placing the OH vertically and along [010]. It matches the one previously identified with machine-learned force fields.[7]

For the water-dosed surfaces, ten starting configurations with the $H_2O$ molecule lying flat on the surface on different surface sites and with different orientations were relaxed. Five of them converged to the lowest-energy configuration shown in Fig. 4c and Fig. S4d, the same structure found in Ref. 6.

Additional details about the β cut

Figures S4e–g show the relaxed models for the β cut. To create the β cut (Fig. S4e), twice as many bonds need to be broken compared to the α cut. As a result, the surface energy of the β cut is significantly larger (Fig. S4h). The subsurface K ions tend to float to the surface to lower its energy when the slab thickness is insufficient. The effect is absent with large-enough slab thicknesses (as mentioned in the Methods, all calculations reported in this work have been performed with 16-layers-thick slabs, which does not induce the floating of the K ions).

Similar to the α cut, $H_2O$ readily dissociates on the β cut (adsorption energy of −3.3 eV/$H_2O$). Here, two (instead of one) $H_2O$ molecules per u.c. are needed to saturate the surface (Fig. S4f). Hence, the full hydroxylation provides a larger energy gain compared to the α cut and an overall lower surface energy (Fig. S4h), consistent with previous findings.[7] The most favored configuration corresponds to $H_2O$ molecules donating their protons to the Al-backbonded surface

O atom and their OH groups to the dangling Si ions. As argued in the main text, the hydroxylated β cut will not be observed experimentally even though it is theoretically preferred at 0 K. A partial coverage (1 $H_2O$/u.c.) was also tried, but it was unfavorable compared to the fully hydroxylated surface.

For completeness, it was also investigated how additional $H_2O$ adsorbs on the hydroxylated β cut – even though, as argued before, this cut was not found to result from cleaving. The most stable configuration is shown in Fig. S4g (adsorption energy of −0.9 eV).

The phase diagram in Fig. S4a summarizes the results obtained for the dry and hydroxylated α and β cuts. It plots their surface energies as a function of the water chemical potential $\mu_{H_2O}$ as defined in the Methods Section in the main text. The dry α cut is more stable than the β cut at all conditions because the β cut requires breaking more bonds. If there is enough water available, both cuts will become hydroxylated. Under UHV-compatible conditions and in the ambient atmosphere, the hydroxylated α cut is more stable. The hydroxylated β cut may become stable under liquid conditions.

Additional details about the α cut

To cleave microcline (001) at the α plane and retain a polarity-compensated surface, 50% of the O atoms lying at the same height as the K ions must be removed. These O atoms are not equivalent, as they are bound to either Si or Al. Hence, different ways exist to cleave the surface. Figure S5 shows the relaxed DFT models obtained by cleaving the slab in three ways and the corresponding surface energies. The lowest-energy structure (used throughout the main text and consistent with the bulk structure used in previous studies[6]) is obtained by exclusively cleaving Al-O bonds and retaining Si-O bonds at the surface (Fig. S5a). The surface energy is significantly larger when the surface is obtained by cleaving Si-O bonds and maintaining Al-O bonds (Fig. S5c). Figure S5b (intermediate surface energy) was obtained by cleaving both Al-O and Si-O bonds. These results are as expected when considering the higher formal charge of the Si atoms (4+) compared with Al (3+).

Performance of r²SCAN and r²SCAN-D3 functionals compared

As shown by Fig. S4i, the optimized r²SCAN and r²SCAN-D3 lattice parameters are in excellent agreement with the experimental values[8] (maximal deviation of ~0.5% and ~0.4%, respectively). The r²SCAN-D3 functional results in only slightly smaller lattice parameters a and c, leading to a minor underestimation of the optimized equilibrium volume compared to the experimental one. The Grimme D3 corrections only slightly increase the surface energies (<6 meV/Å²) and increase the adsorption energies of $H_2O$ (<0.1 eV/$H_2O$) on the different terminations, see Figure S4h.

**Section S4: Considerations about symmetry**

As mentioned in the main text, microcline has a centrosymmetric crystal structure. This means that the (001) and (00$\overline{1}$) facets, while equal in energy, are not equivalent (Fig. S6a). They differ in the positions of the Al ions relative to the tetrahedral framework (related by mirror symmetry).

The same considerations hold for the hydroxylated surface (Fig. S6b). Because the AFM contrast is dominated by the OH groups attached to the Al ions, images simulated on the (001) vs. (00$\bar{1}$) terminations are also mirror symmetric (Fig. S6c). Mirror-symmetric facets have been observed experimentally on different samples (Fig. S6d). The main text reports images acquired mostly on one type of termination, referred to as (001).

## Section S5: Additional experimental and simulated AFM images

Figure S7 illustrates the sensitivity of the AFM contrast to tip terminations. Figure 7a was acquired with a Cu tip (albeit less sharp than the Cu tip used to image the surface as in Fig. 2d in the main text). Figures 7b, c were acquired with tips modified by the interaction with some point defects present at the microcline surface, likely adsorbates. These so-called 'feldspar-modified' tips produce a similar honeycomb pattern as that seen with the sharpest Cu- or CuO$_x$-terminated tips (see main text). Instead, the slightly blunt Cu tip shows a hexagonal pattern with a poor signal-to-noise ratio (approaching any closer induces inadvertent tip changes). Figures S7d, e show the same area of a (00$\bar{1}$)-oriented sample imaged with two different tips. Note that the images are mirror-symmetric compared to the ones in the other panels, in which the sample exposes its (001) facet (see Section S4 above for details about symmetry considerations). Figure S7d was acquired with a (slightly blunt) Cu-prepared tip. Figure S7e shows an image after an interaction with a point defect at the surface, which produces the most common (honeycomb) contrast observed across several samples. The images were aligned using the surface defects as reference. This Cu tip provides a repulsive contrast at the Al-OH (white triangle) and Si-OH (black triangle) positions.

Figures S8a, b highlight the sensitivity of the imaging contrast to the tip-sample separation. In both, the experimental and simulated AFM images, the honeycomb pattern appears more pronounced when approaching closer.

Figures 8c, d compare experimental and simulated images of the water-dosed microcline surface acquired with a Cu-terminated tip. The water features appear dark (attractive) at all explored tip-sample distances. On the other hand, Figs. 5c, d of the main text show that CO-terminated tips produce a repulsive (bright) contrast on the water species. Such differences are expected. According to the model in Fig. 4c, the adsorbed H$_2$O molecules are arranged with an O atom pointing up. Such an O atom should be imaged in the repulsive regime with O-terminated tips, and in the attractive regime with Cu-terminated tips.

## Section S6: Δf-z curves

Curves of frequency shifts vs. tip-sample distance were acquired on representative features on the cleaved and water-dosed microcline (001) surface (Figs. S9c, f). They are marked with the same color coding in the experimental images of Figs. S9b, e and in the proposed DFT models in Figs. S9a, d.

Before each acquisition, the tip was positioned on the chosen feature and retracted by 1 nm from the acquisition height of the images shown in Figs. S9b, e. The two sets of data were acquired with different tips (oxygen-terminated in the case of the hydroxylated surface and modified by the interaction with water on the water-dosed surface), preventing quantitative comparisons. Nonetheless, they can provide rough indications of the vertical separations between the different surface species. A more quantitative investigation of force-distance interactions would demand the acquisition of numerous curves with different, carefully prepared tips and is beyond the scope of this work.

On the cleaved surface (Fig. S9b), curves were acquired on various features discernible at the surface. Based on the assignments in the main text, these correspond to Si-OH (black), Al-OH (light blue), K (orange), and the background between the protruding species (green). The Si-OH curve displays a clear minimum. A shallower minimum is also present for the Al-OH. The curves on the two hydroxyls have their minimum almost at the same z position, in reasonable agreement with the DFT model that predicts a vertical separation between the two hydroxyls of ≈40 pm (to be precise, the vertical separation between the H atoms in the hydroxyls is 50 pm; the one between the O atoms is 26 pm). The Δf separation of the two minima is ≈16 Hz, indicating a different type of interaction of the two types of hydroxyls with the O-terminated tip.[9]

The minimum of the curve acquired on top of the K ion is shifted by roughly 60 pm in z, in reasonable agreement with the calculated vertical distance of ≈90 pm between the surface K ions and the hydroxyls. The curves acquired on the background display the same long-range interactions as the other curves; they do not reach a minimum before −20 Hz.

On the water-dosed surface (Fig. S9d), curves were acquired on several spots on the background (black) and on the water species protruding over the surface (pink). Minima are observed in both cases and are separated by ≈260 pm in height. This indicates that the water species significantly protrude over the surface, in accordance with the DFT models that predict vertical separations of 125 pm and 235 pm between the protruding $H_2O$ and the surface hydroxyls and K ions, respectively.

### Section S7: Further arguments for the ready hydroxylation of the as-cleaved surface

Figures S10a, b compare AFM images of the microcline (001) surface after cleaving and after dosing 0.2 Langmuir (see the Methods section for the definition of Langmuir) $H_2O$ at 100 K, followed by warm-up to 300 K. Because of the high adsorption energies of the hydroxyls predicted by DFT, one expects an initially completely 'dry' surface to readily hydroxylate when exposed to water at low temperatures, and also to keep the hydroxyls when the sample is warmed to 300 K, where additional molecular water desorbs. The similar appearance of the surface after this treatment compared to the as-cleaved surface suggests that hydroxyls were already present after the cleaving.

Figure S10c displays the partial pressure of water during cleaving, as measured with a quadrupole mass-spectrometer in the UHV chamber. A partial pressure spike up to $2 \times 10^{-7}$ mbar is visible. The total integrated area yields a dose of ≈ 24 L, which is much larger than the nominal dose of 0.32 L required to fully hydroxylate the sample. Moreover, the water vapor pressure measured by the mass-spectrometer will be smaller than the one building close to the sample surface.

## Section S8: Imaging in the presence of surface charges

As for other insulators, the microcline (001) surface displays significant charging after cleaving in UHV. If the charge is not compensated by irradiating with X-rays or by applying a sufficiently large bias voltage (see Methods), imaging the surface is challenging. Electrostatic interactions between the surface and the tip dominate the force interactions at large distances, making it hard to judge when the tip is approached to the sample. Once approached, severe background modulations dominate the contrast, complicating the assessment of the surface's tilt and its appropriate correction. Figure S11 shows an example of a (mildly) charged surface. Bright and dark modulations overlap with the microcline lattice. In more severe cases, the background modulations are so strong that they hinder any atomic resolution.

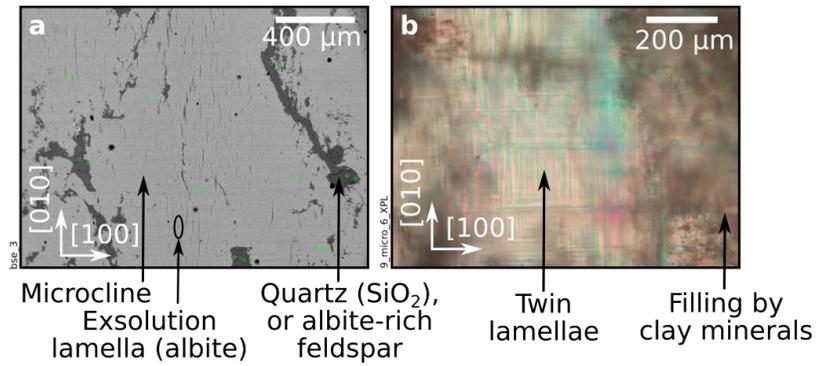

**Figure S1. Characterization of a (001)-oriented section of microcline of ca. 500 μm thickness.** (a) Back scattered electron (BSE) image. Bright grey: microcline, dark grey: albite and/or quartz. (b) Photomicrograph in crossed polarized transmitted light. The sub-horizontally and sub-vertically oriented band-like features are the traces of pericline (sub-horizontal) and albite (sub-vertical) twins.

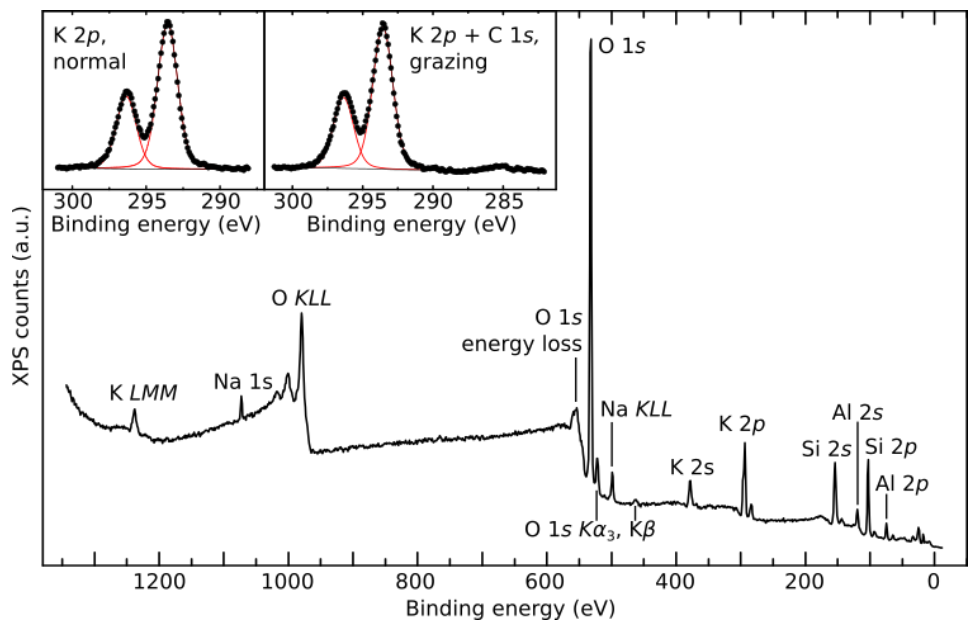

**Figure S2. XPS of UHV-cleaved microcline.** XPS survey (70° grazing emission) of UHV-cleaved microcline (Al Kα, 1486.61 eV, pass energy 60 eV). Inset: K 2*p* and K 2*p* + C 1*s* regions acquired in normal and grazing emission, respectively (pass energy 20 eV; Al Kα₃ satellite removed). All binding energy axes were adjusted to account for charging (see Methods).

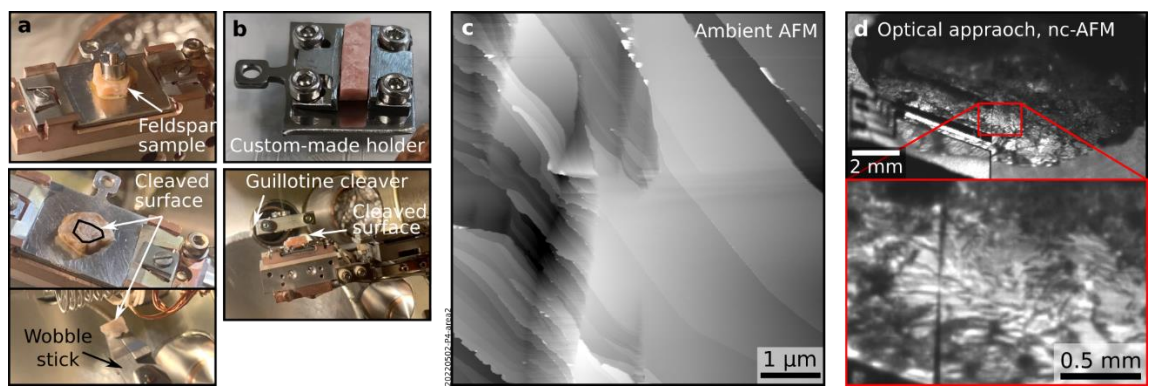

**Figure S3. Cleaving microcline feldspar in UHV.** (a, b) UHV cleaving was performed with either (a) a metal stud glued on the sample or (b) a guillotine-type cleaver. (c) 5 × 5 μm² ambient AFM image of a region of the air-cleaved microcline surface characterized by "small" terraces. (d) Optical approach with the qPlus sensor in UHV.

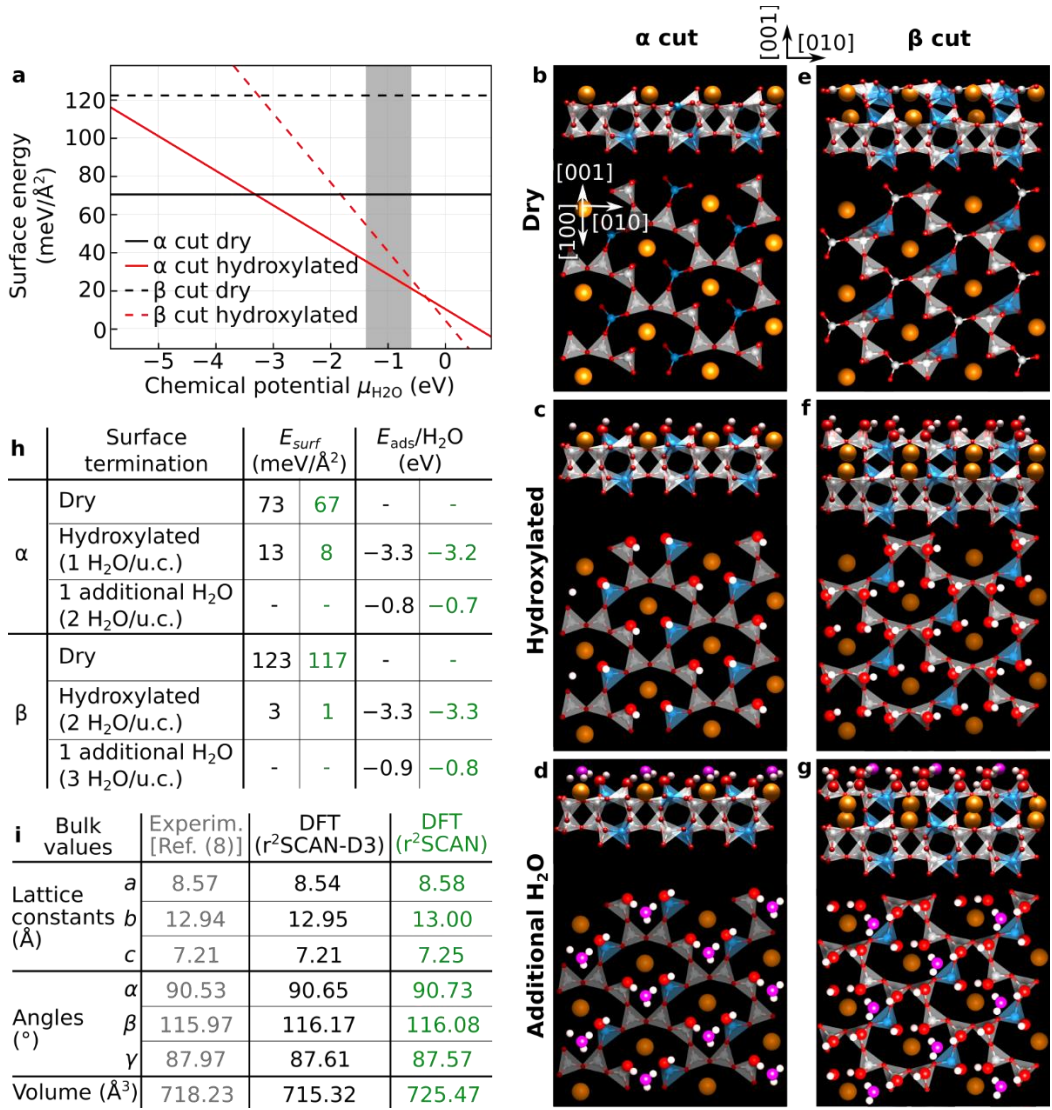

**h**

| Surface termination | $E_{surf}$ (meV/Å²) | | $E_{ads}$/H$_2$O (eV) | |
|---|---|---|---|---|
| **α** Dry | 73 | 67 | - | - |
| Hydroxylated (1 H$_2$O/u.c.) | 13 | 8 | −3.3 | −3.2 |
| 1 additional H$_2$O (2 H$_2$O/u.c.) | - | - | −0.8 | −0.7 |
| **β** Dry | 123 | 117 | - | - |
| Hydroxylated (2 H$_2$O/u.c.) | 3 | 1 | −3.3 | −3.3 |
| 1 additional H$_2$O (3 H$_2$O/u.c.) | - | - | −0.9 | −0.8 |

**i**

| Bulk values | | Experim. [Ref. (8)] | DFT (r²SCAN-D3) | DFT (r²SCAN) |
|---|---|---|---|---|
| Lattice constants (Å) | a | 8.57 | 8.54 | 8.58 |
| | b | 12.94 | 12.95 | 13.00 |
| | c | 7.21 | 7.21 | 7.25 |
| Angles (°) | α | 90.53 | 90.65 | 90.73 |
| | β | 115.97 | 116.17 | 116.08 |
| | γ | 87.97 | 87.61 | 87.57 |
| Volume (Å³) | | 718.23 | 715.32 | 725.47 |

**Figure S4**. **Additional computational results on the (001) microcline surface.** (a) Phase diagram as a function of the water chemical potential (see definition in the Methods section). The grey region identifies the range of pressure between $1 \times 10^{-11}$ mbar and 6 mbar at 300 K. (b−d) Surface structures of relaxed α cuts with (b) no water adsorbed, (c) 1 dissociated H$_2$O/u.c., and (d) 1 H$_2$O molecule/u.c. on top of the hydroxylated surface (reproduced from the main text for reference). (e−g) Relaxed β cuts with (e) no water adsorbed, (f) 2 dissociated H$_2$O/u.c., and (g) 1 H$_2$O/u.c. on top of the hydroxylated surface. (h) Surface energies for $\mu_{H_2O} = 0$ and water adsorption energies on the different terminations obtained with r²SCAN-D3 (black) and the r²SCAN (green) functionals. For the structures containing molecular H$_2$O, differential adsorption energies are given (i.e., the energy of adding a water molecule to the hydroxylated surface). (i) Optimized bulk constants (lattice vectors and angles) and equilibrium volumes obtained with the r²SCAN-D3 (black) and r²SCAN (green) functionals, compared to experiment (8).

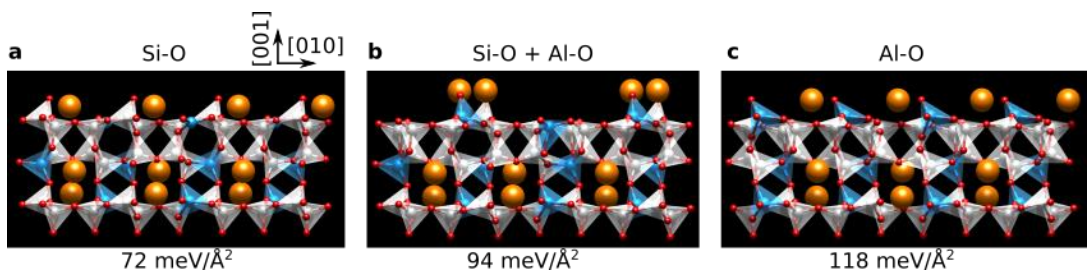

**Figure S5. Relaxed DFT models and surface energies of the α cut of microcline (001) obtained by cleaving different bonds.** (a) Al-O bonds are broken, Si-O bonds remain on the surface. (b) Both Si-O and Al-O bonds are broken. (c) Si-O bonds are broken, Al-O bonds remain.

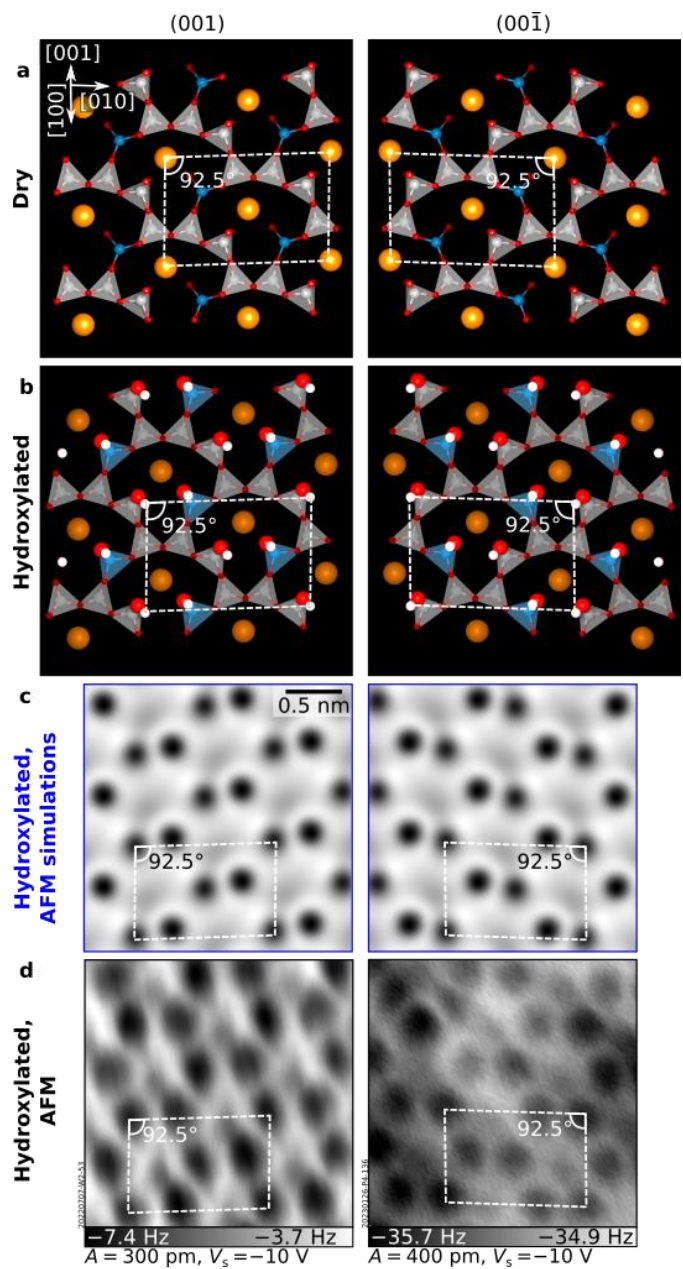

**Figure S6. Considerations about symmetry.** (a) Top views of the (001) and (00$\bar{1}$) terminations of microcline. The structures are mirror-symmetric with respect to the vertical axis. (b, c) Top views and $2.5 \times 2.5$ nm$^2$ AFM simulations of the hydroxylated (001) and (00$\bar{1}$) cuts of microcline. (d) AFM images of two hydroxylated microcline samples imaged with similar tips. The images are mirror symmetric.

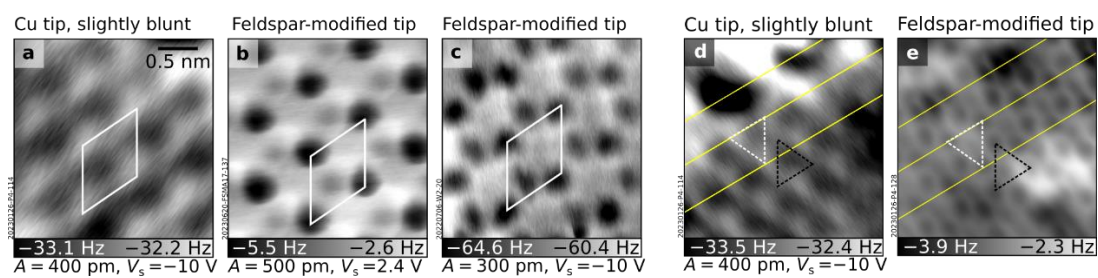

**Figure S7. Imaging with different tip terminations.** (a–c) 2.5 × 2.5 nm² images of UHV-cleaved (001) microcline surfaces measured with the tips specified. The image in panel (a) has been mirrored for displaying purposes – it was acquired on the mirror-symmetric (00$\bar{1}$) orientation, see Section S4. It was acquired with a Cu tip less sharp than the Cu tip used to acquire the image in Fig. 2c of the main text. (d, e) AFM images acquired on the same area of a (00$\bar{1}$)-oriented sample (yellow lines are meant to guide the eye). White and black triangles identify Al-OH and Si-OH, respectively.

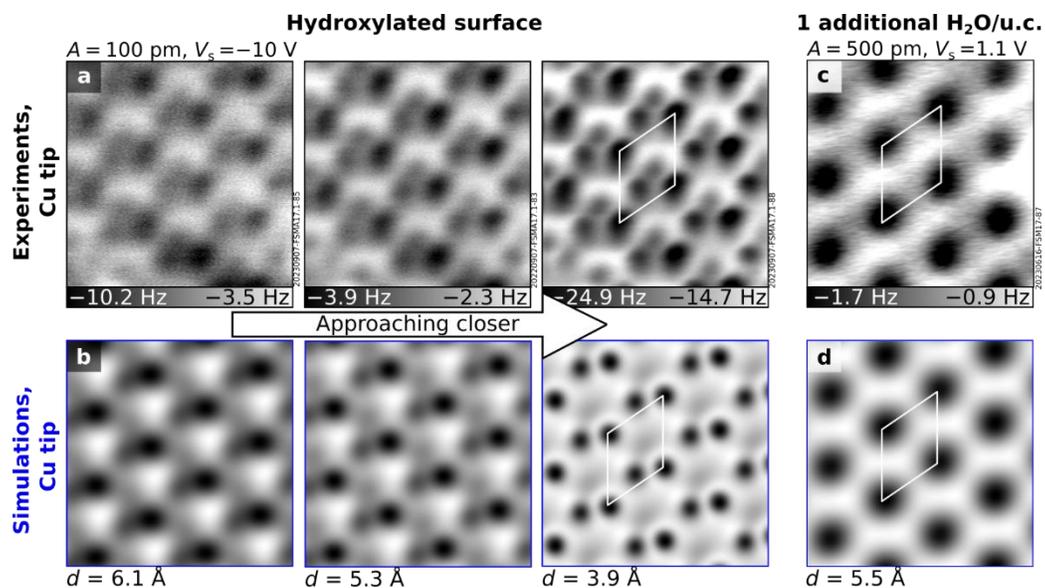

**Figure S8. Experimental and simulated AFM images with Cu tips.** $2.7 \times 2.7$ nm$^2$ experimental (top) and simulated (bottom) AFM images acquired with a Cu tip of the (a, b) hydroxylated, and (c, d) water-dosed microcline (001) α cut. White rhombi identify the primitive unit cells.

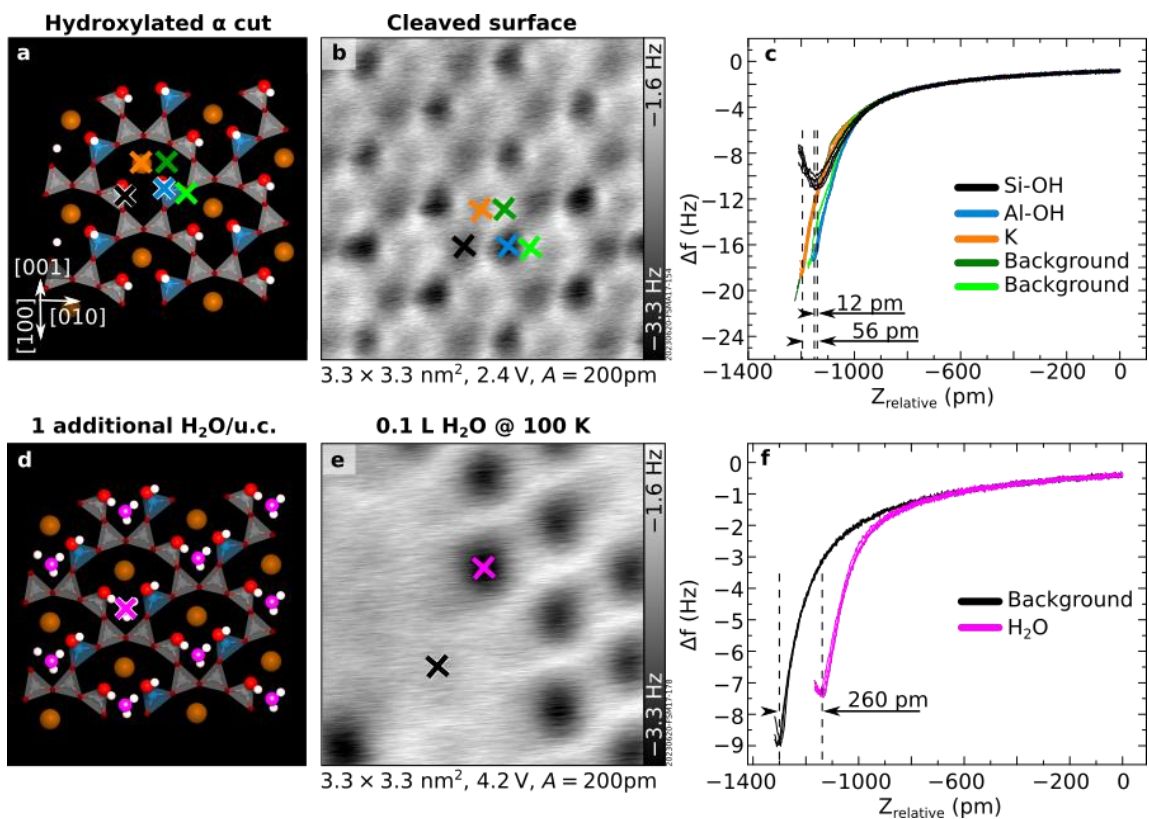

**Figure S9. Force-distance curves**. (a) DFT-relaxed model of the hydroxylated α cut. (c) Δf-z curves acquired on the spots marked in panel (b) with a tip modified by the interaction with a water species. (d) DFT-relaxed model of water of the hydroxylated α cut. (f) Δf-z curves acquired on the spots marked in panel (e) with an oxygen-terminated tip.

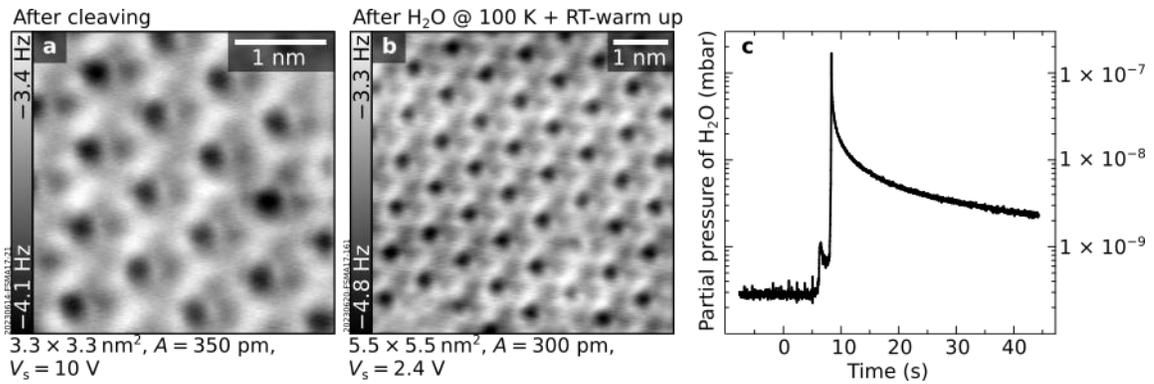

**Figure S10. Ready hydroxylation of microcline in UHV.** AFM images of microcline (001), taken at 4 K, (a) after UHV cleaving and (b) after dosing 0.2 L $H_2O$ at 100 K followed by warming up to room temperature for 20 min. (c) Partial pressure of mass 18 measured by a mass-spectrometer during the UHV cleaving of microcline.

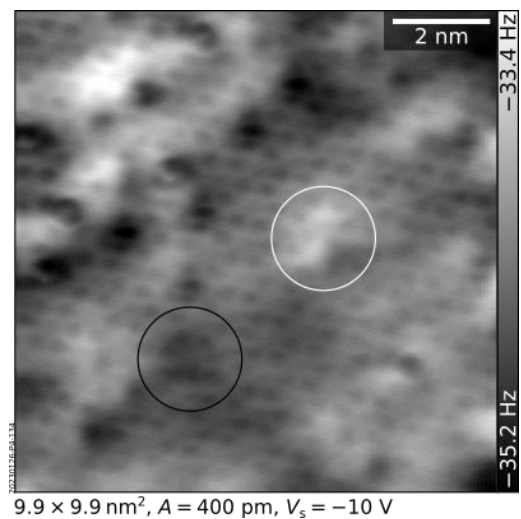

**Figure S11. Appearance of a charged microcline surface in AFM.** Brighter and darker areas (white and black circles) are due to trapped charges.

# Supplementary References